\documentclass[aip,jcp,reprint,twocolumn,showpacs,superscriptaddress]{revtex4-1}

\pdfoutput=1

\usepackage{amsmath,amssymb,gensymb}
\usepackage{bm}
\usepackage{graphicx}

\usepackage{longtable}
\usepackage{upgreek}

\usepackage{color}

\usepackage{ulem}
\newcommand{\rev}[1]{\textcolor{black}{#1}}

\begin{document}

\title{
Thermophoretically induced large-scale deformations around microscopic heat centers 
}

\author{Mate Puljiz}
\email{mate.puljiz@uni-duesseldorf.de}
\affiliation{Institut f\"ur Theoretische Physik II: Weiche Materie, Heinrich-Heine-Universit\"at D\"usseldorf, D-40225 D\"usseldorf, Germany}

\author{Michael Orlishausen}
\email{michael.orlishausen@uni-bayreuth.de}
\affiliation{Physikalisches Institut, Universit\"at Bayreuth, D-95440 Bayreuth, Germany}

\author{Werner K\"ohler}
\email{werner.koehler@uni-bayreuth.de}
\affiliation{Physikalisches Institut, Universit\"at Bayreuth, D-95440 Bayreuth, Germany}

\author{Andreas M.\ Menzel}
\email{menzel@thphy.uni-duesseldorf.de}
\affiliation{Institut f\"ur Theoretische Physik II: Weiche Materie, Heinrich-Heine-Universit\"at D\"usseldorf, D-40225 D\"usseldorf, Germany}

\date{\today}

\begin{abstract}
Selectively heating a microscopic colloidal particle embedded in a soft
elastic matrix is a situation of 
high practical relevance. 
For instance, during hyperthermic cancer treatment, cell tissue surrounding
heated magnetic colloidal particles is destroyed. Experiments on soft elastic
polymeric matrices \rev{suggest} a very long-ranged, \textit{non-decaying} radial
component of the thermophoretically induced displacement fields around the
microscopic heat centers. We theoretically confirm \rev{this conjecture} using a macroscopic hydrodynamic two-fluid description. Both,
thermophoretic and elastic effects are included in this theory. Indeed, we
find that the elasticity of the environment can cause the experimentally
observed large-scale radial displacements in the embedding
matrix. Additional experiments confirm the central role of elasticity.
Finally, a \textit{linearly decaying} radial component of the displacement
field in the experiments \rev{is} attributed to the finite size of \rev{the experimental sample}. \rev{Similar results are obtained from our} theoretical analysis under
\rev{modified} boundary conditions. 
\end{abstract}



\maketitle


\section{Introduction}\label{intro}

When selected inclusions are integrated as fillers into soft polymeric matrices, new composite materials of extended functional application can be created \cite{stankovich2006graphene,balazs2006nanoparticle,thevenot2013magnetic}. Particularly interesting cases concern fillers that can be addressed from outside by external fields. 
One such example are soft magnetic gels \cite{filipcsei2007magnetic,menzel2015tuned,odenbach2016microstructure}, where magnetic colloidal particles \cite{klapp2005dipolar,messina2014self} are embedded into soft elastic environments. 
The result are soft elastic materials, the mechanical properties of which, such as the dynamic behavior \cite{jarkova2003hydrodynamics,bohlius2004macroscopic, filipcsei2007magnetic,tarama2014tunable}, elastic moduli \cite{stepanov2007effect,filipcsei2007magnetic,camp2011modeling,ivaneyko2012effects, borin2013tuning,han2013field,pessot2014structural,menzel2015tuned}, or nonlinear stress-strain behavior \cite{cremer2015tailoring}, 
can be reversibly tuned and switched from outside. 

Another motivation has medical background and mainly concerns cancer treatment. 
Magnetic colloidal particles, possibly loaded with drugs, can be guided to and embedded into cancer tissue using magnetic field gradients \cite{alexiou2006targeting,dobson2006magnetic, tietze2013efficient,zaloga2014development,matuszak2015endothelial}. During hyperthermic cancer treatment \cite{jordan1999magnetic,babincova2001superparamagnetic,lao2004magnetic,hergt2006magnetic}, the colloidal particles are 
heated up. Heat is generated in the particles for instance by rapidly alternating external magnetic fields, as the ongoing remagnetization processes lead to hysteretic losses. When transmitted to the environment, the generated heat can destroy the surrounding cancer cells. 

The nature of the coupling of such colloidal particles to their soft environment has been the subject of ongoing research \cite{messing2011cobalt,weeber2012deformation,gundermann2014investigation,roeder2015magnetic,huang2016buckling}. One point that should be clarified in more detail is the thermal coupling to the environment, which naturally becomes important upon heating the particles. As one aspect, one may ask whether thermophoretic effects \cite{rauch2002diffusion,rauch2003collective} become significant. Thermophoresis describes the observation that temperature gradients induce net forces on molecular and colloidal constituents. 

In recent experiments, this question has been studied on relatively extreme polymeric model variants of the above situation \cite{schwaiger2011transient,schwaiger2013photothermal}. For this purpose, colloidal gold particles were embedded into a polymer solution on the one hand \cite{schwaiger2011transient} and into a strongly entangled 
ultra-high-molecular-weight
polymer network on the other hand \cite{schwaiger2013photothermal}. An individual gold particle could then be selectively heated by external laser irradiation. In this way, \rev{its} surface temperature increased by more than $100$~K and thermophoretic effects could clearly be observed \cite{schwaiger2011transient,schwaiger2013photothermal}. 

For the strongly entangled 
ultra-high-molecular-weight polymer network \cite{schwaiger2013photothermal},
the disentanglement time was huge. 
An overall elastic response of the polymer matrix on the time scale of the
experiment becomes conceivable. In this case, a very long-ranged
outward-oriented radial displacement field around the heated particle was
observed. It was tracked by the displacement of other, non-heated colloidal
particles. 
\rev{These tracers are trapped in the mesh of the transient polymer network. Their motion reflects the local displacement of the polymer matrix. Approximately conserving the local volume, this displacement of the polymer is balanced by a
counter-displacement of the small solvent molecules such that no net flow of
the fluid as a whole is observed. On the one hand, in the experiments, a linear decay of the radial outward displacements around the heated particle was observed \cite{schwaiger2013photothermal}. On the other hand, the associated decay length was of the order of the sample thickness, and the displacements \textit{had} to decay with distance due to the confinement between two rigid plates of high thermal conductivity \cite{schwaiger2013photothermal}. In other words, the
confining boundaries of the sample cavity do not allow an outward
displacement on their inner surfaces \cite{schwaiger2013photothermal}. Therefore, it was conjectured that, remarkably, in an infinite system, the radial outward displacement of the polymer network is in principle constant in magnitude.} That is, the magnitude of the radial displacement
field does not decay with the distance from the center particle. 

So far, this conjecture has not been tested by theoretical analysis. Previous qualitative arguments \cite{schwaiger2013photothermal} did not explicitly include the elasticity of the embedding polymer matrix. Particularly, a theoretical description of the situation would need to cover the possibility for the polymer network to move relatively to the solvent. This enables the relative displacement during the observed thermophoretic motion. 

Here, we perform an according theoretical analysis in \rev{Sec.~\ref{theory_nodecay}}, employing a previously developed macroscopic two-fluid theory \cite{pleiner2004general,pleiner2004generalaip}. We confirm the experimental conjecture of an --- in principle --- non-decaying large-scale radial displacement field around the heated particle. An expression for the magnitude of this displacement field is calculated as a function of the material parameters and the heat rate. Moreover, our analysis demonstrates that elasticity is indeed the likely source for the experimentally observed behavior. 
Our theoretical approach is complemented by experiments on solutions of polystyrene in toluene, see Secs.~\ref{experimental1} and \ref{experimental2}. These experiments are similar to the ones
reported in Ref.~\onlinecite{schwaiger2013photothermal} and contain new data on the heating power
dependence and the time evolution of the thermophoretic displacement of the
transient entanglement network. They highlight the role of elasticity by comparison to less entangled polymer solutions. The experimentally observed linear decay in the radial displacement field is attributed to the finite size of the sample, which is confined between rigid cavity walls. When \rev{confining} boundary conditions are imposed, \rev{a similar} linear decay \rev{is likewise found from our} theoretical results as demonstrated in Sec.~\ref{theory_finite_system}.

\section{Two-fluid description including thermophoretic effects}\label{theory_nodecay}

\rev{To perform our analysis, we employ a recently derived macroscopic two-fluid description \cite{pleiner2004general,pleiner2004generalaip}. For two-component systems, such theories explicitly allow for relative displacements and motion of one component with respect to the other \cite{pleiner2004general,pleiner2004generalaip,pleiner2013active}. In the present case, one component is a simple isotropic liquid playing the role of a solvent, while the other component corresponds to the isotropic elastic polymer network.}

\rev{We are interested in the far-field behavior of the radial displacement field $\mathbf{u}(\mathbf{r})$ of the polymer matrix.
Therefore, we assume weak elastic distortions far from the heated central
particle and thus study a linearized version of the macroscopic two-fluid equations derived in
Refs.~\onlinecite{pleiner2004general,pleiner2004generalaip}. Moreover, we
consider the case of a perfect and permanent elastic network. Due to the long
times of disentanglement for the experimentally employed 
ultra-high-molecular-weight
polymer 
\cite{schwaiger2013photothermal}, 
this represents a reasonable approximation.}

\begin{figure}
\includegraphics[width=5.0cm]{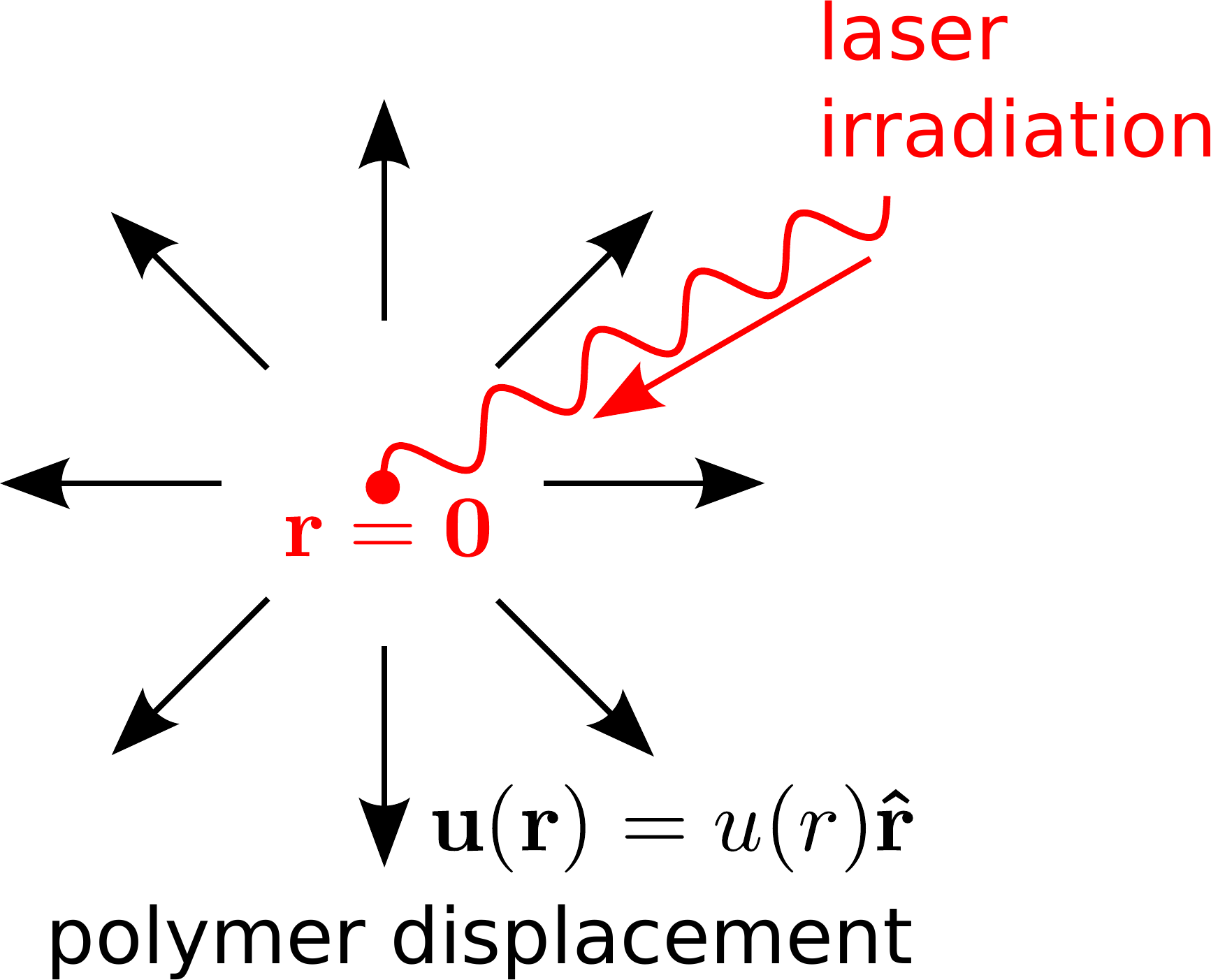}
\caption{\rev{Sketch of the system considered in Sec.~\ref{theory_nodecay}. The red wave represents the laser irradiation that heats the central particle at the origin, resulting in a spherically symmetric displacement field $\mathbf{u}(\mathbf{r})$ (indicated by the black arrows).}}
\label{fig:sketch}
\end{figure}

\rev{In the two-fluid picture, the first fluid of mass density $\rho_1$ represents the solvent, while the second fluid of mass density $\rho_2$ corresponds to the polymer. Assuming constant local density here means that the total mass density $\rho=\rho_1+\rho_2$ be constant. Then the solvent concentration $\phi$ is given by $\phi=\rho_1/\rho$ and the polymer concentration by $1-\phi=\rho_2/\rho$. 
In equilibrium, i.e., in the non-heated state, the solvent concentration is given by a constant value $\phi_{\text{eq}}$. Upon heating, the concentration profile varies by $\delta\phi$, leading to $\phi=\phi_{\text{eq}}+\delta\phi$. The same applies for the entropy density, $s=s_{\text{eq}}+\delta s$.
Moreover, the displacement field $\mathbf{u}$ of the polymer matrix vanishes in equilibrium and becomes non-zero upon heating.
In summary, we characterize the state of our system by the three variables solvent concentration $\phi=\phi_\text{eq}+\delta\phi$, entropy density $s=s_\text{eq}+\delta s$, as well as displacement field $\mathbf{u}$ of the polymer matrix.
The theory in Refs.~\onlinecite{pleiner2004general,pleiner2004generalaip} provides central equations for all these quantities.}

\rev{In the following, we assume a heated colloidal particle in the center at the origin of our coordinate system and perfect spherical symmetry around it. A sketch of the geometry is depicted in Fig.~\ref{fig:sketch}.
As a consequence, we only obtain a radial variation of our variables, i.e.~a concentration field $\phi(r)=\phi_{\text{eq}}+ \delta\phi(r)$, an entropy density $s(r)=s_\text{eq}+\delta s(r)$, as well as a displacement field of the polymer matrix,
\begin{equation}\label{disp}
	\mathbf{u}(r)={}u(r)\mathbf{\hat{r}},
\end{equation}
see Fig.~\ref{fig:sketch}.
Here, $\mathbf{\hat{r}}$ is the radial unit vector in spherical coordinates. From now on, the dependence on $r$ is omitted for briefness of our notation.} 

\rev{The central heat source is considered as point-like. This appears to be a reasonable simplification as the embedded particles in the experiment are of colloidal size and as we use a macroscopic continuum theory for the characterization of the set-up. As will be illustrated in more detail below, the colloidal particles are of approximately $100\,$nm in diameter, while separation distances of about $10-100\,\upmu$m were evaluated, which leads to a separation of length scales of at least two orders of magnitude.
We mentioned above that, in the experimental samples, additional non-heated colloidal particles are embedded that serve as tracer particles for the displacement field \cite{schwaiger2013photothermal}. Their possible influence on the overall behavior is likewise ignored.}

\rev{Due to the strongly entangled state of the polymer matrix, we assume that a final stationary state is reached under steady heating, which will be corroborated by the experimental results below. Therefore, in the following, all time derivatives and macroscopic velocities are equated to zero. This strongly simplifies the situation under investigation. Furthermore, as already mentioned above, we consider the composed system to  keep its local density --- commonly a reasonable assumption for polymeric gels.
As a final prerequisite, we need to introduce the linearized strain tensor \cite{landau1986elasticity}
\begin{equation}\label{lstrain}
	U_{ij}={}\frac{1}{2}(\nabla_i u_j+\nabla_j u_i).
\end{equation}
Inserting Eq.~(\ref{disp}), we obtain
\begin{equation}\label{strain}
	U_{ij}={}u'\hat{r}_i\hat{r}_j+\frac{u}{r}(\hat{\vartheta}_i\hat{\vartheta}_j+\hat{\varphi}_i\hat{\varphi}_j)
\end{equation}
in spherical coordinates with the unit vectors $\mathbf{\hat{r}}$, $\bm{\hat{\vartheta}}$, and $\bm{\hat{\varphi}}$. Moreover, $u'=\partial u(r)/\partial r$. Only the diagonal elements are non-zero.}

\rev{Under all these assumptions, we can now adopt the macroscopic two-fluid equations derived in Ref.~\onlinecite{pleiner2004general} to our situation, before we evaluate them.
We start with the macroscopic equation for the concentration field [Eq.~(120) in Ref.~\onlinecite{pleiner2004general}], which in the stationary limit and for isotropic systems becomes
\begin{equation}\label{conc}
		d\nabla^2(\mu_1-\bar{\mu}_2)+\phi(1-\phi)d^{(T)}\nabla^2 T ={} 0.
\end{equation}
\noindent Here, $d$ is the diffusion coefficient and $d^{(T)}$ denotes the thermodiffusion coefficient.
The field $T$ describes the temperature profile, while $\mu_1$ and $\bar{\mu}_2$ are the chemical potential and effective chemical potential for the mass densities $\rho_1$ and $\rho_2$, respectively.
The first term of Eq.~(\ref{conc}) corresponds to the stationary part of the diffusion equation for a concentration variation $\delta\phi$. This is revealed, when $\delta(\mu_1-\bar{\mu}_2)$ is expressed in our variables, which leads to a contribution $\sim\delta\phi$ [see Eq.~(\ref{deltamu}) below].
In our case, due to thermophoretic effects included by the second term $\sim\nabla^2 T$, nontrivial concentration profiles may arise due to the influence of spatially varying temperature fields.
Here, in the stationary limit, conventional diffusion given by the first term in Eq.~(\ref{conc}) can be balanced by thermodiffusion, described by the second term.}

\rev{Similarly, the} equation for the entropy density [Eq.~(123) in Ref.~\onlinecite{pleiner2004general}] \rev{in the stationary limit for isotropic systems} reads
\rev{\begin{equation}
\kappa\nabla^2 T + \frac{\rho_1\rho_2}{\rho}d^{(T)}\nabla^2(\mu_1-\bar{\mu}_2) ={}-q\delta_D(\mathbf{r}),\label{entr}
\end{equation}
\noindent with $\kappa$ the thermal conductivity. On} its right-hand side, Eq.~(\ref{entr}) contains the source term due to the external heating of the particle at the origin.
$\delta_D(\mathbf{r})$ denotes the Dirac delta function, \rev{treating the heated particle as point-like,} and $q$ sets the external heat rate.
\rev{Without the second term on the left-hand side, Eq.~(\ref{entr}) describes the stationary part of the conventional heat equation with a point-like heat source.
The second term on the left-hand side then includes variations in the temperature profile due to thermophoretic effects.}

Multiplying Eq.~(\ref{conc}) by $\kappa/(d^{(T)}\phi(1-\phi))$ and subtracting the result from Eq.~(\ref{entr}), the \rev{$\nabla^2 T$-terms} are eliminated and we obtain after linearization
\begin{equation}\label{nabla_deltamu_A}
	A\nabla^2\delta(\mu_1-\bar{\mu}_2)={}-q \delta_D(\mathbf{r}),
\end{equation}
with
\begin{equation}
	A={}\rho d^{(T)}\phi_{\text{eq}}(1-\phi_{\text{eq}})-\frac{d\kappa}{ d^{(T)}}\frac{1}{\phi_{\text{eq}}(1-\phi_{\text{eq}})}.
\end{equation}
Therefore,
\begin{equation}\label{deltamu_A}
	\delta(\mu_1-\bar{\mu}_2)={}\frac{q}{4\pi A}\frac{1}{r}+\tilde{A},
\end{equation}
\noindent with a constant $\tilde{A}$. As a function of $U_{ij}$, $\delta s$, and $\delta\phi$, variations $\delta(\mu_1-\bar{\mu}_2)$ can be expressed as [see Eq.~(131) in Ref.~\onlinecite{pleiner2004general}]
\begin{equation}\label{deltamu}
	\delta(\mu_1-\bar{\mu}_2)={}\frac{1}{\kappa_u}U_{kk} + \frac{1}{\rho\alpha_\phi}\delta s+\frac{1}{\rho\kappa_\phi}\delta\phi,
\end{equation}
with the expansion coefficient $\alpha_\phi$ and the compressibilities $\kappa_{\phi,u}$.
The trace of the strain tensor, see Eq.~(\ref{strain}), reads
\begin{equation}\label{trace}
	U_{kk}={}u'+\frac{2u}{r}.
\end{equation}
Inserting Eqs.~(\ref{deltamu_A}) and (\ref{trace}) into Eq.~(\ref{deltamu}), we obtain
\begin{equation}\label{GL1}
	\frac{q}{4\pi A}\frac{1}{ r}+\tilde{A} ={} \frac{1}{\kappa_u} \left(u'+\frac{2u}{r}\right) +\frac{1}{\rho\alpha_\phi}\delta s+ \frac{1}{\rho\kappa_\phi}\delta\phi.
\end{equation}

In a subsequent step, we return to Eq.~(\ref{conc}). There, we \rev{insert} Eq.~(\ref{nabla_deltamu_A}) and find
\begin{equation}
	B\nabla^2 T ={} -q\delta_D(\mathbf{r}),\label{laplace_T}
\end{equation}
with
\begin{equation}
	B ={} -A\frac{d^{(T)}\phi_{\text{eq}}(1-\phi_{\text{eq}})}{d}.
\end{equation}
Integration of Eq.~(\ref{laplace_T}) yields
\begin{equation}
	\delta T = \frac{q}{4\pi B}\frac{1}{r}+\tilde{T}\label{T},
\end{equation}
with a constant $\tilde{T}$.
Moreover, an expansion of the temperature variation is given by [Eq.~(130) in Ref.~\onlinecite{pleiner2004general}]
\begin{equation}\label{TEMP}
	\delta T={} \frac{1}{\alpha_3}U_{kk} + \frac{T_{\text{eq}}}{C_V}\delta s+\frac{1}{\alpha_\phi}\delta\phi  ,
\end{equation}
with $\alpha_3$ an expansion coefficient, $C_V$ the specific heat, and $T_{\text{eq}}$ the equilibrium temperature. Inserting Eq.~(\ref{T}) into Eq.~(\ref{TEMP}), we obtain
\begin{equation}\label{GL2}
	\frac{q}{4\pi B}\frac{1}{r}+\tilde{T} ={} \frac{1}{\alpha_3}\left(u'+\frac{2u}{r}\right)+\frac{T_{\text{eq}}}{C_V}\delta s+\frac{1}{\alpha_\phi}\delta\phi.
\end{equation}

Next, from the equation for the relative velocity between the two components [Eq.~(128) in Ref.~\onlinecite{pleiner2004general}], we find in the stationary limit
\begin{equation}\label{musigma}
	\nabla_i(\mu_1-\bar{\mu}_2)+\nabla_j\frac{1}{\rho_2}\sigma_{ij}={}0,
\end{equation}
with $\sigma_{ij}$ the stress tensor.
\rev{Summation over repeated indices is implied.}
After linearization, this equation may be rewritten as
\begin{equation}
	\nabla_j\big[\sigma_{ij} + \rho(1-\phi_{\text{eq}})\delta(\mu_1-\bar{\mu}_2)\delta_{ij} \big] ={}0.\label{sigma_gamma_grundgleichung}
\end{equation}
Thus, we may express $\sigma_{ij}$ as
\begin{equation}
	\sigma_{ij} ={} -\rho(1-\phi_{\text{eq}})\delta(\mu_1-\bar{\mu}_2)\delta_{ij}+\Gamma_{ij},\label{sigma_integriert}
\end{equation}
where we have introduced a diagonal tensor $\Gamma_{ij}$ that satisfies $\nabla_j\Gamma_{ij}=0$. The diagonal form of $\Gamma_{ij}$ is justified by the expression for the stress tensor $\sigma_{ij}$ as a function of $U_{ij}$, $\delta s$, and $\delta\phi$ [see Eqs.~(90) and (91) in Ref.~\onlinecite{pleiner2004general}], which takes the form
\begin{eqnarray}
	\sigma_{ij}&=&{}  c_{\text{tr}}\left(U_{ij}-\frac{1}{3}\delta_{ij}U_{kk}\right)+\frac{1}{3}c_{\text{l}}\delta_{ij}U_{kk} \nonumber\\
	 &&{}+\frac{1}{3\alpha_3}\delta_{ij}\delta s+\frac{1}{3\rho\kappa_u}\delta_{ij}\delta\phi.\label{stress}
\end{eqnarray}
In this expression, $c_{\text{tr}}$ and $c_{\text{l}}$ are the transversal and longitudinal elastic moduli, respectively. 
From Eq.~(\ref{stress}) together with Eq.~(\ref{strain}) it follows that 
$\sigma_{\vartheta\vartheta}=\sigma_{\varphi\varphi}$. An analogous relation thus applies for $\Gamma_{ij}$. $\nabla_j\Gamma_{ij}=0$ therefore yields the additional condition
\begin{equation}
	\Gamma_{rr}'+2\frac{\Gamma_{rr}}{r}-2\frac{\Gamma_{\vartheta\vartheta}}{r} ={} 0.\label{GL3}
\end{equation}
Introducing Eq.~(\ref{sigma_integriert}) into Eq.~(\ref{stress}), we obtain two different equations, namely the
$\mathbf{\hat{r}\hat{r}}$ component,
\begin{eqnarray}\label{GL4}
	\frac{q}{4\pi C}\frac{1}{r}+ \tilde{C} +\Gamma_{rr} &={} & \frac{2}{3}c_{
		\text{tr}}\left(u'-\frac{u}{r}\right)  +\frac{1}{3}c_{\text{l}} \left(u'+\frac{2u}{r}\right) \notag\\
	&{}&+ \frac{1}{3\alpha_3}\delta s+\frac{1}{3\rho\kappa_u}\delta\phi,
\end{eqnarray}
and the $\bm{\hat{\vartheta}\hat{\vartheta}}$ or $\bm{\hat{\varphi}\hat{\varphi}}$ component,
\begin{eqnarray}\label{GL5}
	\frac{q}{4\pi C}\frac{1}{r}+ \tilde{C} +\Gamma_{\vartheta\vartheta} &={} & \frac{1}{3}c_{
		\text{tr}}\left(\frac{u}{r} - u'\right)  +\frac{1}{3}c_{\text{l}} \left(u'+\frac{2u}{r}\right) \notag\\
	&{}&+ \frac{1}{3\alpha_3}\delta s+\frac{1}{3\rho\kappa_u}\delta\phi,
\end{eqnarray}
with the constants
\begin{equation}
	 C = -\frac{A}{\rho(1-\phi_{\text{eq}})}, \qquad \tilde{C}= -\rho(1-\phi_{\text{eq}})\tilde{A}.\label{C_schlange}
\end{equation}

In the macroscopic two-fluid description of Ref.~\onlinecite{pleiner2004general}, there is another equation resulting for the elastic degrees of freedom [Eq.~(124) in Ref.~\onlinecite{pleiner2004general}]. However, in the stationary case and for our assumptions, this equation is satisfied identically (we consider a perfectly elastic network, which on the considered time scale does not disentangle; moreover, vacancy diffusion \cite{martin1972unified} is neglected). Apart from that, the assumption of overall constant density sets the pressure field [see Eqs.~(127) and (129) in Ref.~\onlinecite{pleiner2004general}].

Therefore, our remaining task is to solve the system of Eqs.~(\ref{GL1}), (\ref{GL2}), (\ref{GL3}), (\ref{GL4}), and (\ref{GL5}). We use as an ansatz for $u$, $\delta s$, $\delta\phi$, $\Gamma_{rr}$, and $\Gamma_{\vartheta\vartheta}$, respectively, a power series in the radial distance~$r$,
\begin{eqnarray}
	u &={} &\sum\limits_{n=-\infty}^{\infty} u_n r^n,\label{ansatz_u}\\	
	\delta s &={} &\sum\limits_{n=-\infty}^{\infty} \delta s_n r^n,\label{ansatz_s}\\
	\delta\phi &={} &\sum\limits_{n=-\infty}^{\infty} \delta\phi_n r^n,\label{ansatz_phi}\\
	\Gamma_{rr} &={} &\sum\limits_{n=-\infty}^{\infty} \Gamma_{rr,n}r^n,\\
	\Gamma_{\vartheta\vartheta} &={} &\sum\limits_{n=-\infty}^{\infty} \Gamma_{\vartheta\vartheta,n}r^n.\label{ansatz_gamma_theta}
\end{eqnarray}

Inserting this ansatz into Eqs.~(\ref{GL1}), (\ref{GL2}), (\ref{GL3}), (\ref{GL4}), as well as (\ref{GL5}), \rev{we obtain equations that can be sorted by order in $r^n$.
Solving the equations for each order of $r^n$ separately, we find the system of equations}
\begin{eqnarray}
	 0&={}&(n+2) \frac{1}{\kappa_u} u_n + \frac{1}{\rho\alpha_\phi}  \delta s_{n-1} + \frac{1}{\rho\kappa_\phi} \delta\phi_{n-1}, \label{alle_n_A}\\
	 0&={}&(n+2) \frac{1}{\alpha_3} u_n + \frac{T_{\text{eq}}}{C_V}  \delta s_{n-1} + \frac{1}{\alpha_\phi} \delta\phi_{n-1}, \label{alle_n_B}\\
	 0&={}&(n+1)\Gamma_{rr,n-1}-2\Gamma_{\vartheta\vartheta,n-1},\label{alle_n_Gamma}\\
	 0&={}& \frac{1}{3}\big(2(n-1)c_{\text{tr}} +(n+2)c_{\text{l}} \big)u_n\notag\\
	 	 &{}& + \frac{1}{3\alpha_3}\delta s_{n-1} + \frac{1}{3\rho\kappa_u}\delta\phi_{n-1}  -\Gamma_{rr,n-1} \label{alle_n_sigma1},\\
	 0 &={}& \frac{1}{3}\big((1-n)c_{\text{tr}} + (n+2)c_{\text{l}}\big) u_n \notag\\
	 	 &{} &+ \frac{1}{3\alpha_3}\delta s_{n-1} + \frac{1}{3\rho\kappa_u}\delta\phi_{n-1} -\Gamma_{\vartheta\vartheta,n-1} \label{alle_n_sigma2},
\end{eqnarray}
where $n\in\mathbb{Z}\setminus\{0,1\}$. 
We note that there is no coupling between sets $\{u_n, \delta s_{n-1}, \delta\phi_{n-1},\Gamma_{rr,n-1},\Gamma_{\vartheta\vartheta,n-1}\}$ of different $n\in\mathbb{Z}\setminus\{0,1\}$. Moreover, the heat source $q$, which here is responsible for driving the system out of equilibrium, does not appear in the above equations. 
Thus it follows from the equations that \rev{our} expansion coefficients for $n\in\mathbb{Z}\setminus\{0,1\}$ keep their equilibrium values equal to zero.

An exception, for which no strict statement is obtained, are the coefficients $u_{-2}$, $\Gamma_{rr,-3}$, and $\Gamma_{\vartheta\vartheta,-3}$. $u_{-2}$ cancels in Eqs.~(\ref{alle_n_A}) and (\ref{alle_n_B}), while we obtain from the remaining equations:
\begin{equation}
	u_{-2} ={} -\frac{\Gamma_{rr,-3}}{2c_{\text{tr}}}, \qquad \Gamma_{\vartheta\vartheta,-3} ={} -\frac{\Gamma_{rr,-3}}{2}. \label{eq:redundance}
\end{equation}
Thus, the precise magnitude of the coefficient $u_{-2}$, strictly speaking, remains unassigned in the present framework. It may be determined by the microscopic processes and properties on the heated particle surface, which are not captured by our macroscopic theory \cite{pleiner1996pattern}. Yet, this coefficient is insignificant for our purpose as we are interested in the far-field behavior, while the contribution $\sim r^{-2}$ decays rapidly in the far-field displacement. The far-field displacement is dominated by a \rev{long-ranged} leading order identified in the following.

For $n=1$, we find from Eqs.~(\ref{GL1}), (\ref{GL2}), (\ref{GL3}), (\ref{GL4}), and (\ref{GL5}) the expressions
\begin{eqnarray}
	 \tilde{A}&={}& \frac{3}{\kappa_u} u_1 + \frac{1}{\rho\alpha_\phi}  \delta s_{0} + \frac{1}{\rho\kappa_\phi} \delta \phi_{0},\label{u_1_a} \\
	 \tilde{T}&={}& \frac{3}{\alpha_3} u_1 + \frac{T_{\text{eq}}}{C_V}  \delta s_{0} + \frac{1}{\alpha_\phi} \delta \phi_{0}, \\
	 0&={}&\Gamma_{rr,0}-\Gamma_{\vartheta\vartheta,0}, \label{redundant}\\
	 \tilde{C}&={}& c_{\text{l}} u_1  + \frac{1}{3\alpha_3}\delta s_{0} + \frac{1}{3\rho\kappa_u}\delta \phi_{0}  -\Gamma_{rr,0}.\label{u_1_e}
\end{eqnarray}
One of the two relations in Eqs.~(\ref{alle_n_sigma1}) and (\ref{alle_n_sigma2}) becomes redundant due to Eq.~(\ref{redundant}).
Thus, additional conditions are necessary to fix the magnitude of the expansion coefficients  $u_1$, $\delta s_{0}$, $\delta \phi_{0}$, and $\Gamma_{rr,0}$.
At $r\rightarrow\infty$ the magnitude of displacement should not diverge. This sets $u_1=0$. In addition to that, we assume that for $r\rightarrow\infty$ the temperature and the (effective) chemical potentials keep their equilibrium values, e.g., via coupling to external heat and substance reservoirs. From Eqs.~(\ref{deltamu_A}), (\ref{T}), and (\ref{C_schlange}) 
it then follows that $\tilde{A}=\tilde{T}=\tilde{C}=0$ and $\delta s_0=\delta\phi_0=\Gamma_{rr,0}=\Gamma_{\vartheta\vartheta,0}=0$. 

Finally, we find for the remaining order $n=0$ \rev{the following system of linear equations:}
\begin{eqnarray}
	 \frac{q}{4\pi A}&={}& \frac{2}{\kappa_u} u_0 + \frac{1}{\rho\alpha_\phi}\delta  s_{-1} + \frac{1}{\rho\kappa_\phi} \delta\phi_{-1}, \label{n_0_gl_a}\\
	 \frac{q}{4\pi B}&={}& \frac{2}{\alpha_3} u_0 + \frac{T_{\text{eq}}}{C_V} \delta s_{-1} + \frac{1}{\alpha_\phi} \delta\phi_{-1},\\
	 \frac{q}{4\pi C}&={}& \frac{2}{3}(c_{\text{l}}-c_{\text{tr}}) u_0+ \frac{1}{3\alpha_3}\delta s_{-1} \notag\\
	 	 	 &{}& + \frac{1}{3\rho\kappa_u}\delta\phi_{-1}  -\Gamma_{rr,-1}, \\
	 \frac{q}{4\pi C} &={}& \frac{1}{3}(c_{\text{tr}} + 2c_{\text{l}}) u_0 + \frac{1}{3\alpha_3} \delta s_{-1} \notag\\
	 	 	 &{} &+ \frac{1}{3\rho\kappa_u}\delta\phi_{-1} -\Gamma_{\vartheta\vartheta,-1}, \\
	 0&={}&\Gamma_{rr,-1}-2\Gamma_{\vartheta\vartheta,-1}.\label{n_0_gl_e}
\end{eqnarray}
Here, the heat source $q$ directly enters and drives the system out of equilibrium.
\rev{These equations can be solved explicitly, which leads to expressions for the coefficients $u_0$, $\delta s_{-1}$, $\delta\phi_{-1}$, $\Gamma_{rr,-1}$, and $\Gamma_{\vartheta\vartheta,-1}$.
They are rather lengthy and therefore are listed in the appendix. The important point is that we have now obtained an analytical solution of the macroscopic equations, which reveals the overall long-ranged response of the system.}

\begin{figure}
\includegraphics[width=7.5cm]{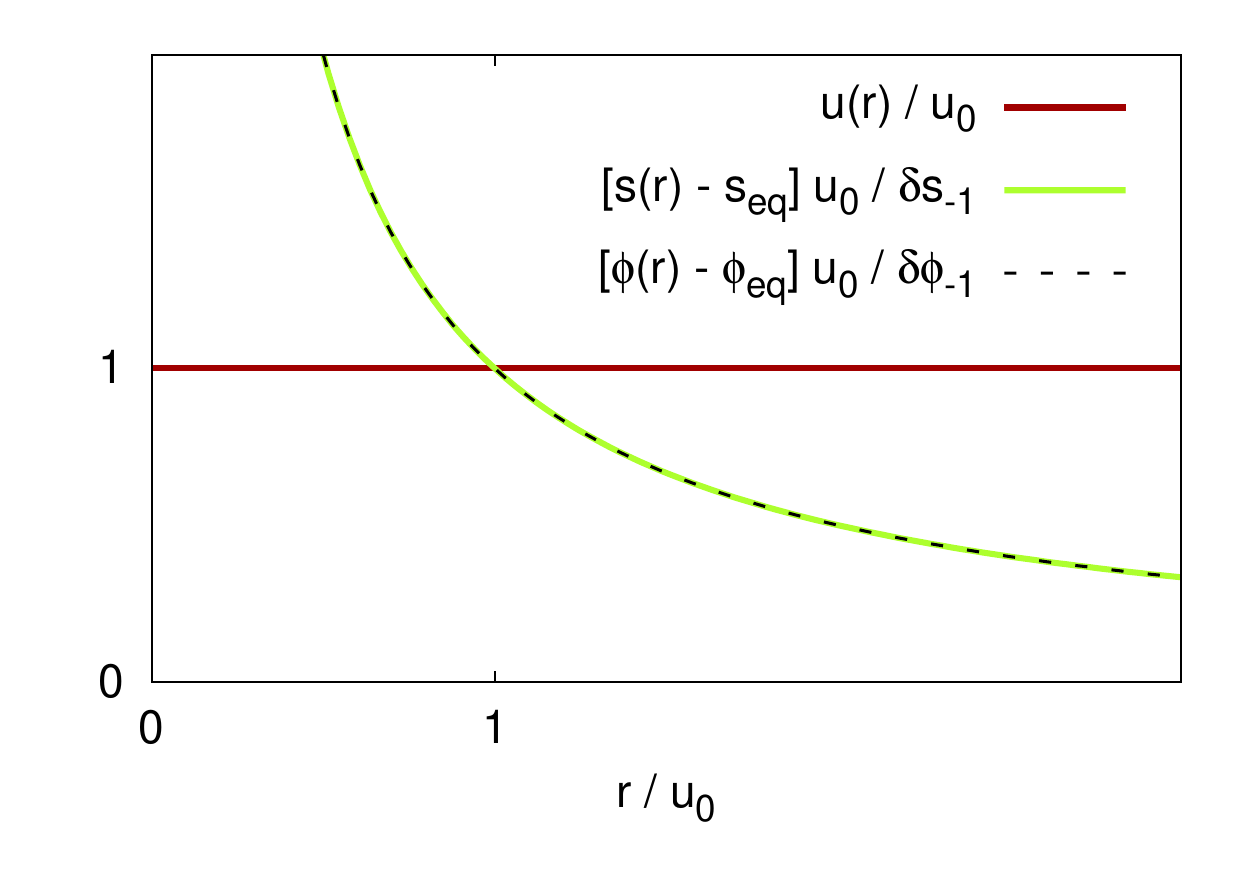}
\caption{\rev{Plots of the rescaled leading-order deviations from static equilibrium for $u(r)$, $s(r)$, and $\phi(r)$ as functions of the distance $r$ from the heated center, see Eqs.~(\ref{s_final})--(\ref{u_final}). The rapidly decaying $r^{-2}$ term in Eq.~(\ref{u_final}) is neglected. Overall, our macroscopic solutions highlight the long-ranged far-field behavior, see also the discussion in the text.}}
\label{fig:th_infinite}
\end{figure}

In \rev{summary}, the \rev{results for the} entropy density and the concentration \rev{profile are obtained as}
\begin{eqnarray}
	 s(r) &=&{}  s_{\text{eq}}+\frac{\delta s_{-1}}{r},\label{s_final} \\ 
	 \phi(r) &=&{} \phi_{\text{eq}}+\frac{\delta\phi_{-1}}{r}. \label{phi_final}
\end{eqnarray}
\rev{Very illustratively, in these expressions the $1/r$ dependence of the entropy and concentration deviations can be viewed as a direct consequence of the spherical symmetry with the heated source in the center. From the heat equation, see Eq.~(\ref{entr}), we find the $1/r$ temperature profile given by Eq.~(\ref{T}). This well-known result survives the thermophoretic coupling to the concentration field described by Eqs.~(\ref{conc}) and (\ref{entr}). With the $1/r$ temperature profile at hand, already the expansion of the temperature fluctuations in our variables in Eq.~(\ref{TEMP}) suggests the $1/r$ dependence of the entropy and concentration deviations found in Eqs.~(\ref{s_final}) and (\ref{phi_final}), see also Fig.~\ref{fig:th_infinite}.}

\rev{Thus, the irradiated heat drives the system out of equilibrium, which is directly reflected by the modified entropy density and, via thermophoretic coupling, by the modified concentration field.
Changing the local concentration in our spherically symmetric situation is only possible by radial displacement of the elastic polymer network. However, radial displacements imply strains, see also Eq.~(\ref{trace}), which directly contribute to the chemical potential and temperature variations, see Eqs.~(\ref{deltamu}) and (\ref{TEMP}). Naturally, opposing stresses arise from these distortions, see Eq.~(\ref{stress}), which counteract the effect. Overall,}
the general solution for the \rev{radial} displacement field reads
\begin{equation}\label{u_final}
	u(r)={}u_0+\frac{u_{-2}}{r^2},
\end{equation}
\rev{and contains a \textit{non-decaying} component $u_0$.}
As discussed above, the precise magnitude of the coefficient $u_{-2}$ remains unassigned in the present framework. Yet, its impact decays rapidly as a function of distance and does not influence the leading-order \rev{far-field} behavior connected with the coefficient $u_0$.
\rev{The leading-order deviations of $u(r)$, $s(r)$, and $\phi(r)$ from static equilibrium are illustrated in Fig.~\ref{fig:th_infinite} as functions of $r$.}

\rev{At first glance, one might wonder how a non-decaying radial outward displacement of constant magnitude is energetically possible throughout the system. After all, the amount of displaced material diverges with increasing distance $r$ from the center. For the moment, let us consider spherical shells of thickness $\mathrm{d}r$ around the center. Then, the number of displaced volume elements on each shell diverges with its volume as $\sim r^2\mathrm{d}r$. These volume elements are stretched along the directions tangential to the shell surface during radial outward displacement, which costs elastic energy. However, the corresponding local strain energy density\cite{landau1986elasticity} following via the strain given by Eq.~(\ref{trace}) decreases as $\sim 1/r^2$. Overall, we obtain a constant strain energy $\sim\mathrm{d}r$ per shell. Illustratively, the reason for this result is the following. There are many more strained volume elements on the shells for large $r$. However, the curvature of these shells decreases with increasing $r$. As a consequence, the outward displacement with increasing $r$ locally more and more resembles a rigid outward translation of the locally nearly flat parts of the shell surfaces. Rigid translations do not cost elastic energy. Still, in total, the non-decaying radial displacement field requires an infinite input of strain energy in our infinite system. Yet, it is provided via the infinite input of heat energy (already the $1/r$ temperature profile resulting for the simple decoupled static heat equation corresponds to an infinite energy input in an infinite system).}

Overall, \rev{with our central result in Eq.~(\ref{u_final})}, we here confirmed in a two-fluid approach that it is indeed the elastic response of one of the two components that can induce \rev{a} \textit{non-decaying} radial \rev{outward} displacement \rev{$u_0$}, as conjectured in Ref.~\onlinecite{schwaiger2013photothermal}. Moreover, we derived an expression for the magnitude of $u_0$, \rev{see Eq.~(\ref{sol_u}) in the appendix}, as a function of the macroscopic material parameters, \rev{where, however,} several numerical values of these macroscopic parameters are not known at present. 

%
%

\section{Experimental setup and sample preparation}\label{experimental1}

To further elaborate on the role of elasticity during the temperature-induced
deformation process and to further study the influence of a finite system
size, we have performed additional experimental investigations of the
thermophoretic effect around laser-heated gold nanoparticles (GNPs) in
transient polymer networks.

\begin{figure}
\includegraphics[width=8.5cm]{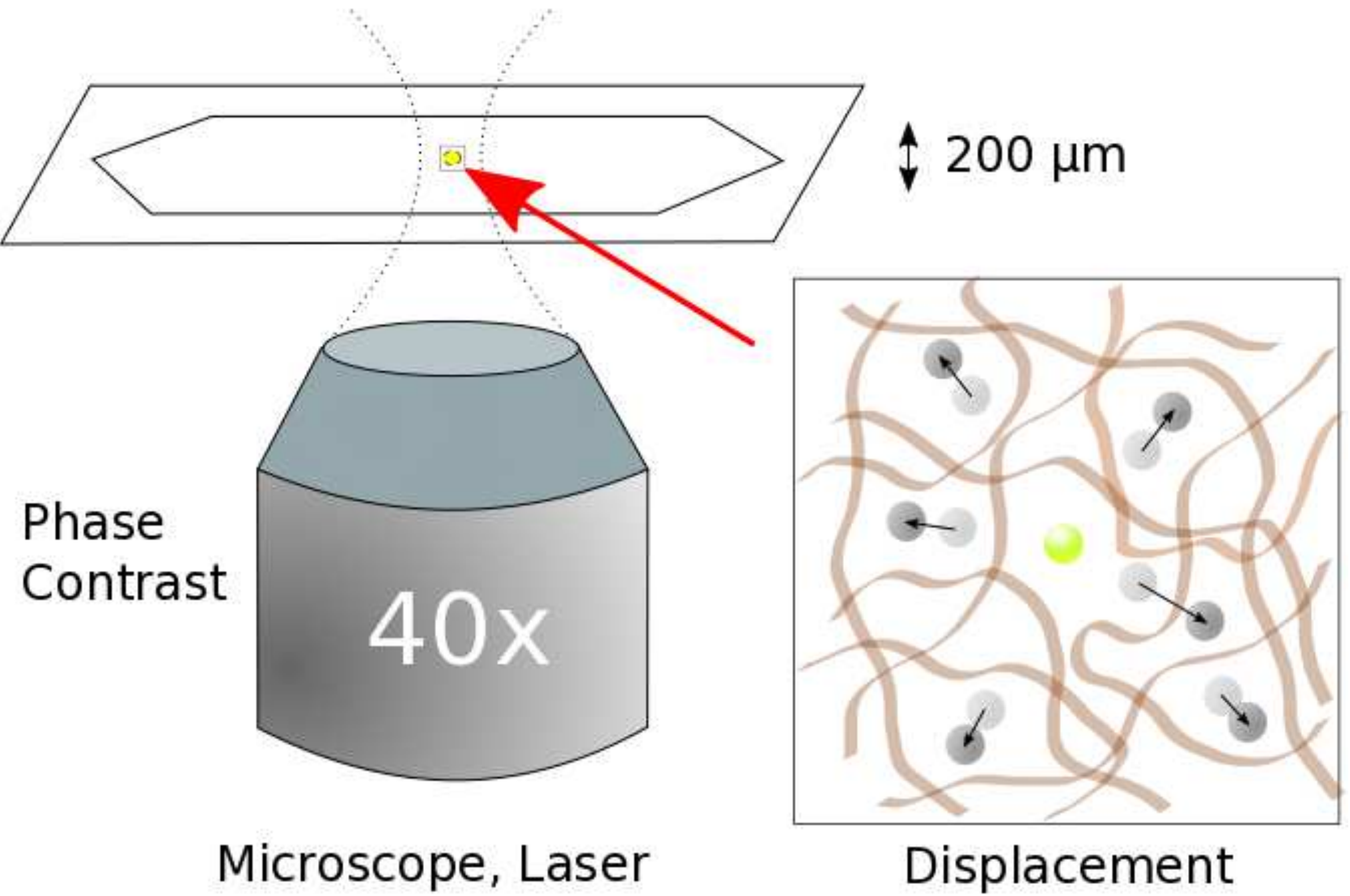}
\caption{Experimental set-up to visualize thermophoretically induced
displacements. A focused laser beam heats a single gold nanoparticle (GNP),
indicated in yellow. Other GNPs (grey) are trapped as tracers in an entangled
polymer network and follow its displacement. The sample consists of a thin
film of about $200\,\upmu\text{m}$ in thickness, confined between rigid
cuvette walls. }
\label{fig:setup}
\end{figure}

GNPs (BBInternational) of approximately $100\,\text{nm}$ diameter show a
strong optical absorption in their plasmon resonance.  They serve as tunable
heat sources that can individually be addressed by a tightly focused laser
beam ($\lambda = 525\,\text{nm}$, Coherent Verdi V5), see Fig.~\ref{fig:setup} for an illustration of our set-up.
The non-heated GNPs, which are randomly distributed over the sample, serve as
tracers to visualize the deformation of the transient network.  Their
displacement is monitored with an optical microscope (Olympus IX71) equipped
with an sCMOS camera (Tucsen Discovery MH15). Phase contrast imaging is used
for the observation of smaller particles.

%
%

Samples were prepared by first dissolving polystyrene (PS) of different
molecular weight ($M_w = 16\,800\,\text{kg}\,\text{mol}^{-1}$, $M_w =
500\,\text{kg}\,\text{mol}^{-1}$, PSS) in toluene (Sigma Aldrich, 99.9\%) at a
weight concentration of 6\%, yielding a highly entangled polymer solution. 
It is then slowly stirred with 20 rpm for at least 10 days. A small droplet
of the polymer solution is filled into a detachable cuvette (Hellma 106-QS).
After evaporation of the solvent, the thin left-over polymer film is overlayed
with a droplet of the aqueous GNP dispersion. The GNPs are given approximately
10 minutes to sediment and attach to the polymer layer, before the remaining
GNP dispersion is removed by flushing with deionized water (Millipore).
After drying, the remaining volume is filled with polymer
solution. The cuvette is closed and sufficient time is given for the
polymer solution to redissolve the lower polymer layer and to homogenize,
resulting in a transient entangled polymer network with randomly embedded
GNPs.
The final polymer concentration in the cuvette is approximately
twice the one of the stock solution. 
Sedimentation of the GNPs is very slow and can be compensated by occasionally
turning over the cuvette.

Several samples were examined 
using laser powers between $2\,\text{mW}$ and $20\,\text{mW}$. Due to the
small cross section of the GNPs and the much wider laser focus, the absorbed
laser power is several orders of magnitude smaller.  Particle displacement was 
determined 
by fitting a two dimensional Gaussian function, which allows for a position
accuracy of roughly $\pm$ $50\,\text{nm}$.

\section{Measurements}\label{experimental2}

\begin{figure}
\includegraphics[width=8.5cm]{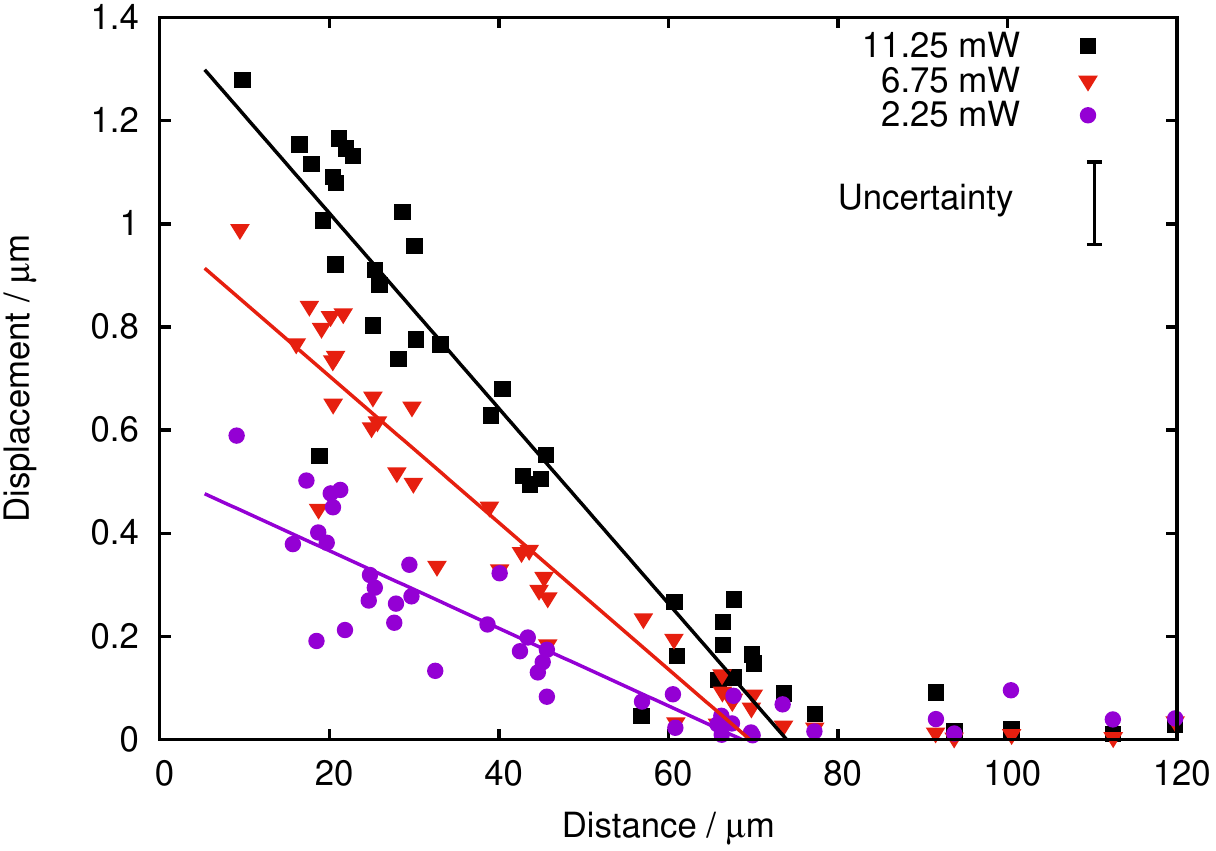}
\caption{Distance-dependent radial displacements of GNPs for different
heating powers as indicated on the top right. Linear fits illustrate an approximately linear decay with radial distance from the heated center. All data points were acquired at a time of $2\,\text{s}$ after switching on the heating. 
(The initial concentration of the polymer was $c=0.06$.)}
\label{fig:displacement}
\end{figure}

%
\rev{We have recorded the displacements of the tracer particles in the plane parallel to the cell windows, which we define as the $xy$ plane. The optical axis was perpendicular to this plane.}
The displacements of GNPs within a distance of $10\,\upmu\text{m}$ to
$120\,\upmu\text{m}$ to the heated center have been tracked and are plotted
in Fig.~\ref{fig:displacement}.  
These displacements 
show a behavior similar to the 
 ones reported by Schwaiger et al.\ \cite{schwaiger2013photothermal},
however, with a somewhat extended range.
Most likely, this is a consequence of
the higher polymer concentration of the stock solution ($c=0.06$ as compared
to $c=0.03$), but no systematic study of the concentration dependence has been
carried out so far.

As indicated in Fig.~\ref{fig:displacement}, the amplitude of the long-ranged
radial displacements shows an approximately linear decay
with the distance from the heated center.  This linear decay is
attributed to the finite size of the sample, which is confined by rigid
cuvette walls \rev{of sevenfold higher thermal conductivity} that enforce vanishing displacements at the cell boundaries
\cite{schwaiger2013photothermal}. 
Thus, our experimental data do not contradict our theoretical results in
Sec.~\ref{theory_nodecay}. There, an infinitely extended system has been
considered, while a rigidly confined system will be discussed below in
Sec.~\ref{theory_finite_system}. Apart from that, a sublinear dependence of
the displacement 
amplitude 
on the heating power was observed. 

\begin{figure}
\includegraphics[width=8.5cm]{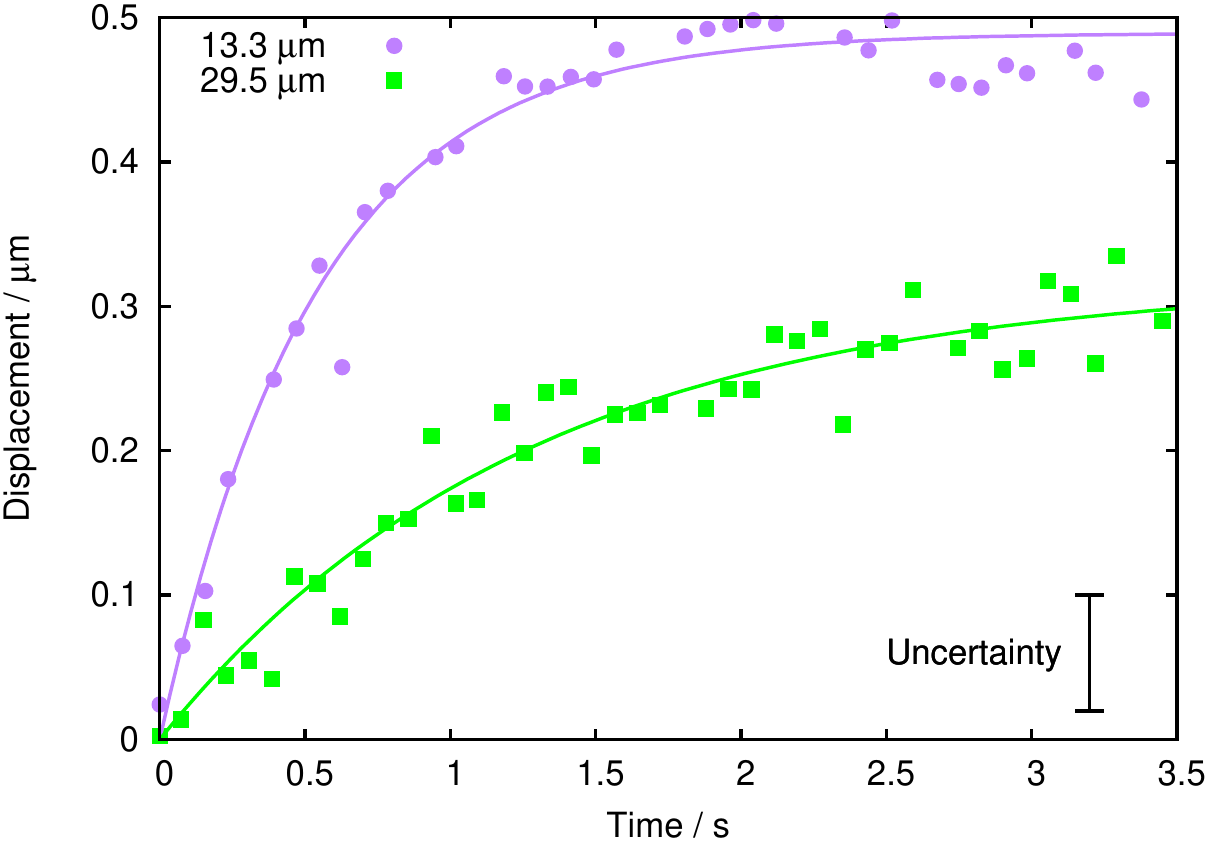}
\caption{Time dependence of the displacement of two different GNPs at
different distances from the heated center as indicated on the top left. The fits were performed using an exponential with a single relaxation time, yielding $(0.54\pm0.04)\,\text{s}$ and $(1.27\pm0.19)\,\text{s}$ for the two distances, 
respectively \cite{supplemental}. 
A laser power of $2.25\,\text{mW}$ was applied.}
\label{fig:time_evolution}
\end{figure}

All displacements in Fig.~\ref{fig:displacement} were measured at a fixed time
of $2\,\text{s}$ after turning on the heating.  We plot the time evolution of
the displacement for two different examples of lower laser power in
Fig.~\ref{fig:time_evolution}. 
As can be inferred from Fig.~\ref{fig:time_evolution}, 
the characteristic time scale 
to reach a \mbox{(quasi-)}stationary displacement increases with radial
distance from the heated GNP. The fits 
correspond to 
an exponential with a single relaxation time. As we can
see, relaxation is to a 
large
extent completed at our measurement time of
$2\,\text{s}$. 
We have checked that the linear decay in Fig.~\ref{fig:displacement} is still
observed when plotting the extrapolated asymptotic displacements instead of
the transient values after $2\,\text{s}$ \cite{supplemental}.

Finally, we underline the role of the elastic response of the strongly
entangled polymer network. In accord with our theoretical considerations in
Sec.~\ref{theory_nodecay}, it is the most likely source for the observed
long-ranged displacements. For this purpose, we performed identical
measurements using, however, polymers of significantly lower molecular weight
of $M_w =
500\,\text{kg}\,\text{mol}^{-1}$ at the same concentration. This hardly entangled polymer solution does not feature a long-time elastic response.  

\begin{figure}
\includegraphics[width=8.5cm]{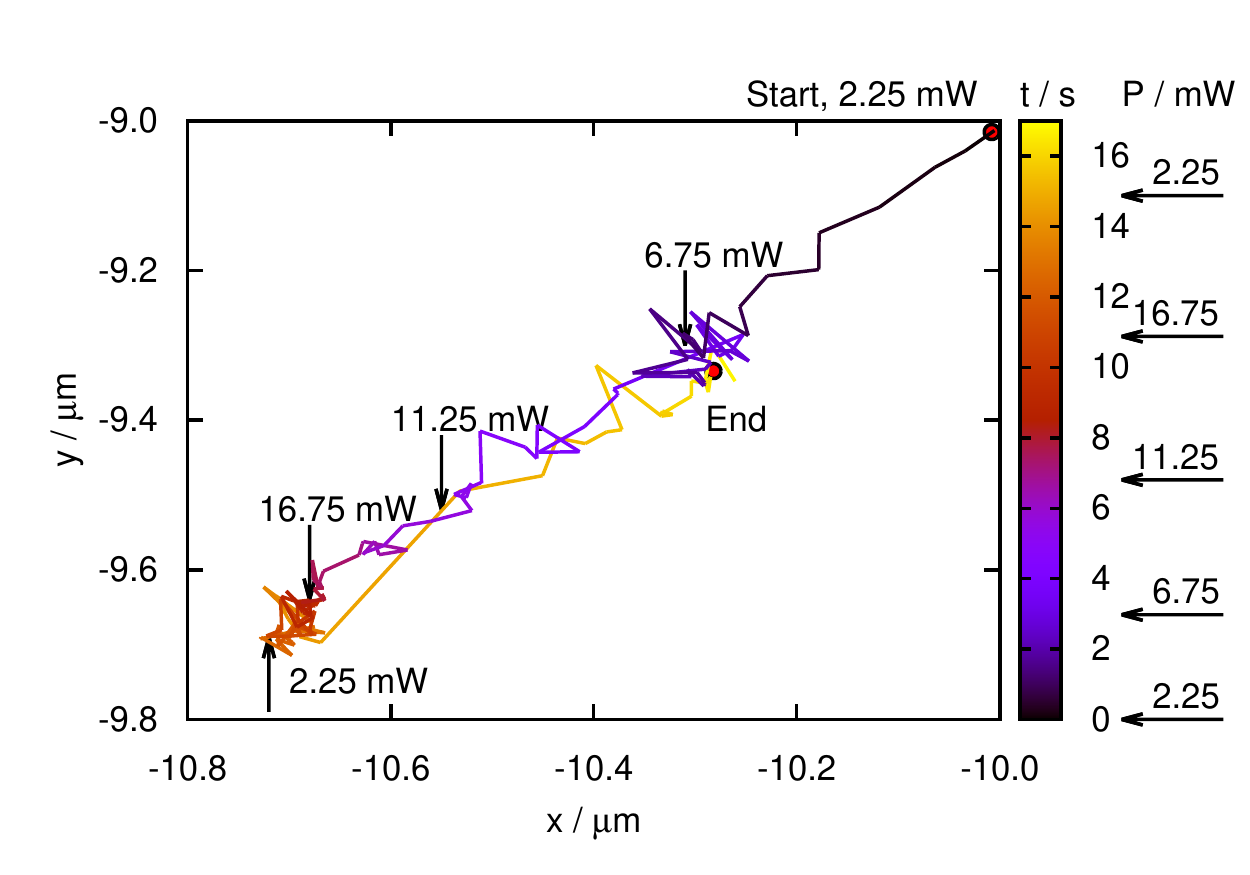}
\caption{Example trajectory of a tracer GNP at an initial distance of approximately
$15\,\upmu\text{m}$ from the heated center in a $c = 0.06$ solution
of molecular weight $M_w = 16\,800\,\text{kg}\,\text{mol}^{-1}$.
\rev{The heated center is located at the origin.}
A protocol of sudden stepwise increase and subsequent decrease of the laser power $0\,\text{mW}\rightarrow2.25\,\text{mW}\rightarrow6.75\,\text{mW}\rightarrow11.25\,\text{mW}\rightarrow16.75\,\text{mW}\rightarrow2.25\,\text{mW}$ is performed. The numbers on the trajectory mark switches to the new (indicated) laser power, from where relaxation to the new steady state occurs. 
The coincidence of the point of switching $2.25\,\text{mW}\rightarrow6.75\,\text{mW}$ and of the end point at $2.25\,\text{mW}$ highlights the reversible character of the displacements.}
\label{fig:trajectory_high}
\end{figure}

\begin{figure}
\includegraphics[width=8.5cm]{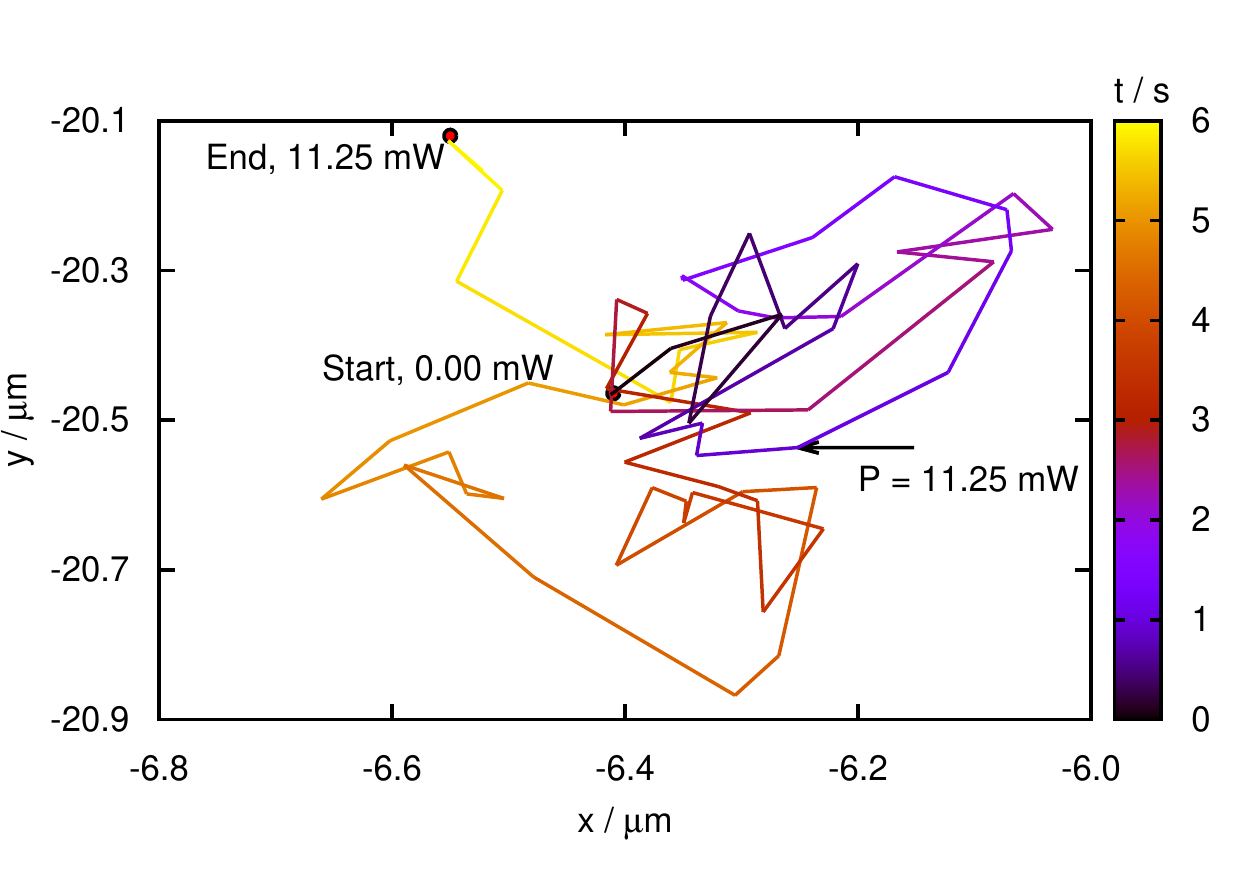}
\caption{Example trajectory of a tracer GNP at an initial distance of approximately
$15\,\upmu\text{m}$ from the heated center in a $c = 0.06$ solution
of $M_w =500\,\text{kg}\,\text{mol}^{-1}$.
\rev{The heated center is located at the origin.}
Here, the GNP is subject to visible
Brownian motion due to the relatively low viscosity compared to the highly entangled
solution in Fig.~\ref{fig:trajectory_high}. The laser is switched on only at the point marked by the arrow at about
$1.5\,\text{s}$ into the recorded trajectory. Turning on the laser does not have any visible effect. 
} \label{fig:trajectory_low}
\end{figure}

Typical example trajectories of tracer GNPs in the high- and low-molecular-weight samples relatively close to the heated center are shown in Figs.~\ref{fig:trajectory_high} and \ref{fig:trajectory_low}, respectively. 
During a heating protocol of switching on and off the laser, the GNP in the high-molecular-weight sample is undergoing 
a directed and reversible displacement. In contrast to that, the displacements of the GNP in the lower-molecular-weight sample are dominated by Brownian motion. In this lower-molecular-weight case, Brownian motion is more pronounced because of the lower viscosity of the solution. Other than for the entangled system, switching on the laser does not lead to correlated displacements.

\section{Theoretical description of rigidly confined systems}\label{theory_finite_system}

Our generalized theoretical derivation in Sec.~\ref{theory_nodecay} 
has been 
performed for an infinitely extended system, while in the experiments, by
construction, samples of finite size were investigated. Most importantly, the
experimental samples are confined by rigid cuvette walls and an approximately
linear decay in the outward displacement field was observed, see
Fig.~\ref{fig:displacement}. We explain in the following that \rev{similar results can be found within} our theoretical \rev{framework}, \rev{when modified}
boundary conditions \rev{are applied}. 

To address finite size, we consider the system to be confined within a spherical shell of effective radius $R$ \cite{schwaiger2013photothermal}. This \rev{geometry} keeps the problem within our analytical framework of Sec.~\ref{theory_nodecay}. Assuming the shell to be rigid, the outward displacement must vanish on its surface: 
\begin{equation}\label{finite_u_bc}
	u(r=R) ={} 0.
\end{equation}
\rev{Naturally, the rigid spherical shell does not exactly match the experimental configuration of rigid confinement between two parallel plates. However, the presence of the confining cuvette windows in the experiments sets a dominant length scale for the overall response of the system. The cuvette walls have a sevenfold higher thermal conductivity, which tends to bend the direction of the heat fluxes towards the plate surfaces. As a consequence, both geometries, the experimental situation and the rigid sphere, are strongly influenced by the one dominating length scale set by the confinement.}

Returning to Sec.~\ref{theory_nodecay}, we find that our results and conclusions from Eqs.~(\ref{alle_n_A})--(\ref{eq:redundance})\rev{, (\ref{n_0_gl_a})--(\ref{n_0_gl_e}), and (\ref{sol_u})--(\ref{sol_abb})}, particularly our expression for $u_0$, remain unchanged \rev{upon imposing Eq.~(\ref{finite_u_bc})}. Modifications only concern our conclusions from Eqs.~(\ref{u_1_a})--(\ref{u_1_e}). Previously, the coefficient $u_1$ only vanished from our requirement that $u(r\rightarrow\infty)$ \rev{should not diverge}. Under the modified boundary condition Eq.~(\ref{finite_u_bc}), we instead obtain
\begin{equation}
u_1 = -\frac{u_0}{R},
\end{equation}
where we now directly neglected a possible --- but within the present framework undetermined --- coefficient $u_{-2}$. Its contribution $\sim r^{-2}$ vanishes rapidly with increasing distance from the heated center. 
Altogether, this leads to 
\begin{equation}\label{u_finite}
	u(r) ={} u_0\left(1-\frac{r}{R}\right) 
\end{equation}
and thus \rev{coincides with} the linear decay observed in the experiments.


Let us briefly discuss the implications for the other variables in
Eqs.~(\ref{u_1_a})--(\ref{u_1_e}). For instance, one may maintain the
equilibrium temperature $T(r=R)=T_{\text{eq}}$ on the system boundary by
coupling to an external heat bath. This sets the constant $\tilde{T}=-q/4\pi B
R$. Assigning a value to $\tilde{A}$ is difficult as it, in principle,
requires knowledge about chemical interaction details with the shell surface.
In an actually closed spherical cavity, the overall confined material should
be conserved. The constant $\tilde{A}$ can be used to tune $\delta\phi_0$ to
achieve this overall goal within the framework of our solution.  It should be
recalled, however, that in our experiment the system is not enclosed in a
sphere and exchange of material in the lateral direction is possible. 

\section{Conclusions}

In this work, we have analyzed the situation of 
an elastic two-fluid system subjected to thermophoretic effects. The two-fluid environment consists of an elastic matrix suspended in a fluid solvent. These two components can be displaced relatively with respect to each other by externally heating a colloidal particle embedded in the system center. Recent experiments reported here and in Ref.~\onlinecite{schwaiger2013photothermal} suggest a long-ranged \textit{non-decaying} component of the resulting thermophoretically induced radial displacement field \rev{in an idealized infinite system}. The elastic response of the polymer matrix seems to play a central role in these observations. 
Using a macroscopic two-fluid description \cite{pleiner2004general,pleiner2004generalaip}, our theoretical analysis confirms this experimental conjecture \rev{in the linearized regime}. 
Moreover, for rigidly confined systems, a \textit{linear decay} in the radial displacement field is experimentally observed. \rev{Similar results are found from} theoretical considerations for rigidly bounded spherical systems. 

\rev{We should recall at this point that our analysis was performed for small deviations from equilibrium only, using a macroscopic theory. That is, we lost several couplings between different variables during the linearization of the equations. For example, Eqs.~(\ref{conc}) and (\ref{entr}) initially contain nonlinear couplings to the concentration deviations. In dynamic situations, further nonlinearities are present due to the convective terms but also due to nonlinear contributions to the reversible stress, see Ref.~\onlinecite{pleiner2004general}. Predominantly, deviations from the linearized regime are of course obtained close to the heated center. Yet, in the very vicinity of the center, the macroscopic description loses its significance in any case and microscopic processes need to be taken into account. Most immediately, our macroscopic approach serves to characterize the far-field behavior, which was the central scope of our investigation.}

Concerning future investigations, we remark that our present analysis is restricted to the case where a final steady state is attained. As a next step of significantly higher complexity, 
dynamic situations could be addressed. On the one hand, this applies for the dynamic path towards a final steady state. \rev{In nonlinear situations, the final state may even depend on the chosen dynamic path towards it.} On the other hand, a final steady state might not be reached at all \rev{on the considered time scales}, if the polymer matrix is not perfectly elastic but shows disentanglement or vacancy diffusion \cite{martin1972unified,pleiner2004general,pleiner2004generalaip}. Moreover, both theoretically and experimentally, situations of more than one heated particle could be addressed. Finally, the significance of the observed effect in similar situations but different systems should be analyzed. A related example with medical background is biological tissue exposed to hyperthermic cancer treatment \cite{jordan1999magnetic,babincova2001superparamagnetic,lao2004magnetic,hergt2006magnetic}.

\appendix
\section*{Appendix}
\rev{Eqs.~(\ref{n_0_gl_a})--(\ref{n_0_gl_e}) form a system of linear equations that can be solved explicitly for the coefficients $u_0$, $\delta s_{-1}$, $\delta\phi_{-1}$, $\Gamma_{rr,-1}$, and $\Gamma_{\vartheta\vartheta,-1}$.
Here we list the explicit solutions. The resulting expressions read
\begin{widetext}
\begin{eqnarray}
	u_0 &={} & q \alpha_3 \kappa_u
	\bigg(
	\rho d^{(T)} T_\text{eq} \alpha_3\alpha_\phi^2
			\Big(
				-\kappa_\phi + 3\rho \kappa_u(-1+\phi_\text{eq})
			\Big)
			\Big(
				-1+\phi_\text{eq}
			\Big) \phi_\text{eq}
			+ C_V
			\Big(
				-d\rho\alpha_\phi^2 \kappa_u \notag\\
				&{}&- 3\rho^2 d^{(T)}\alpha_3\kappa_u\kappa_\phi (-1+\phi_\text{eq})^2\phi_\text{eq}
				+ \alpha_\phi\kappa_\phi
				\big(
					d\alpha_3 + \rho^2 d^{(T)}\kappa_u(-1+\phi_\text{eq})\phi_\text{eq}
				\big)
			\Big)
	\bigg)
	\bigg/2D, \label{sol_u}\\
	s_{-1} &={} &{} q \rho C_V \alpha_3\alpha_\phi
	\Big(
		-d\alpha_3\alpha_\phi\kappa_\phi 
		- 3\rho d^{(T)}\alpha_\phi \kappa_u^2 \phi_\text{eq} 
		+ 3\rho^2 d^{(T)}\alpha_3\kappa_u\kappa_\phi\phi_\text{eq}
		- d^{(T)}\alpha_\phi\kappa_u\kappa_\phi\phi_\text{eq}\notag\\
				&{}&
		+6\rho d^{(T)}\alpha_\phi\kappa_u^2\phi_\text{eq}^2 -6\rho^2 d^{(T)} \alpha_3\kappa_u\kappa_\phi\phi_\text{eq}^2
		+d^{(T)}\alpha_\phi \kappa_u\kappa_\phi\phi_\text{eq}^2
		-3\rho d^{(T)}\alpha_\phi\kappa_u^2\phi_\text{eq}^3
		+3\rho^2 d^{(T)}\alpha_3\kappa_u\kappa_\phi\phi_\text{eq}^3\notag\\
				&{}&
		+c_\text{l}\alpha_3\kappa_u^2
		\big(
			d\alpha_\phi-\rho d^{(T)}\kappa_\phi(-1+\phi_\text{eq})\phi_\text{eq}
		\big)
		+2 c_\text{tr}\alpha_3\kappa_u^2
		\big(
			d \alpha_\phi - \rho d^{(T)}\kappa_\phi (-1+\phi_\text{eq})\phi_\text{eq}
		\big)
	\Big)
	 \Big/D,\label{sol_s}\\
	\delta\phi_{-1} &={} &{}-q \rho \alpha_\phi \kappa_u \kappa_\phi
	\bigg(
		\rho d^{(T)} T_\text{eq} \alpha_3^2\alpha_\phi
		\Big(
			-c_\text{l}\kappa_u-2c_\text{tr}\kappa_u + 3\rho\big(-1+\phi_\text{eq}\big)
		\Big)
		\Big(-1+\phi_\text{eq}
		\Big)
		\phi_\text{eq}
		+ C_V
		\Big(
			d
			\big(
				c_\text{l}\notag\\
									&{}& + 2c_\text{tr}
			\big)
			\alpha_3^2\kappa_u+\rho d^{(T)}\alpha_\phi\kappa_u
			\big(
				-1+\phi_\text{eq}
			\big)
			\phi_\text{eq}
			-\rho\alpha_3
			\big(
				d\alpha_\phi + 3d^{(T)}\kappa_u (-1+\phi_\text{eq})^2\phi_\text{eq}
			\big)
		\Big)
	\bigg)
	\bigg/D,\label{sol_phi}\\
	\Gamma_{rr,-1} &={} &  q c_\text{tr}\alpha_3 \kappa_u
	\bigg(
		-\rho d^{(T)} T_\text{eq} \alpha_3\alpha_\phi^2
		\Big(
			-\kappa_\phi
			+3\rho\kappa_u
			\big(
				-1+\phi_\text{eq}
			\big)
		\Big)
		\Big(
			-1+\phi_\text{eq}
		\Big)
		\phi_\text{eq}
		+ C_V
		\Big(
			d\rho \alpha_\phi^2\kappa_u\notag\\
			&{}&
			+3\rho^2 d^{(T)}\alpha_3\kappa_u\kappa_\phi
			\big(
				-1+\phi_\text{eq}
			\big)^2
			\phi_\text{eq}
			-\alpha_\phi\kappa_\phi
			\big(
				d\alpha_3 + \rho^2 d^{(T)}\kappa_u
				(-1+\phi_\text{eq})\phi_\text{eq}
			\big)
		\Big)
	\bigg)
	\bigg/D, 
\end{eqnarray}
\begin{eqnarray}
	\Gamma_{\vartheta\vartheta,-1} &={} &\frac{1}{2}\Gamma_{rr,-1},\label{sol_gtt}
\end{eqnarray}
with the abbreviation
\begin{eqnarray}\label{sol_abb}
 D&=&{} 4\pi d^{(T)}
 	\Big(\rho T_{\text{eq}} \alpha_3^2\alpha_\phi^2 \big(c_\text{l} \kappa_u^2  + 2 c_\text{tr} \kappa_u^2 - \kappa_\phi\big) + C_V \kappa_u
 	\big(-\rho \alpha_\phi^2 \kappa_u + (1+\rho^2) \alpha_3\alpha_\phi\kappa_\phi \notag\\
 	&&{}- \rho (c_\text{l} + 2 c_\text{tr} ) \alpha_3^2 \kappa_u \kappa_\phi\big)\Big) (-1+\phi_\text{eq}) \phi_\text{eq} 
 \left(\rho d^{(T)}\phi_{\text{eq}}(1-\phi_{\text{eq}})-\frac{d\kappa}{ d^{(T)}}\frac{1}{\phi_{\text{eq}}(1-\phi_{\text{eq}})}\right).
\end{eqnarray}
\end{widetext}
All coefficients $u_0$, $\delta s_{-1}$, $\delta\phi_{-1}$, $\Gamma_{rr,-1}$, and $\Gamma_{\vartheta\vartheta,-1}$ are directly proportional to the external heat rate set by $q$. Therefore, they directly characterize the deviation of the system from static equilibrium and are non-zero only when the external heat source is switched on.
At present, the numerical values of several of the listed material parameters are not known.}

\begin{acknowledgments}
M.P.\ and A.M.M.\ thank the Deut\-sche For\-schungs\-ge\-mein\-schaft (DFG) for support of this work through the priority program SPP 1681. \\
\end{acknowledgments}


%

\end{document}